\def   \ni {\noindent}

\def   \ssk {\vskip  5truept}

\def   \bsk {\vskip 15truept}
\def   \newpage {\vfill\eject}
\def   \newline {\hfil\break}

\def\ms{M$_{\odot}$}
\def\zs{Z$_{\odot}$}
\def\cs{cm$^{-2}$ s$^{-1}$}
\def\ga{$\gamma$}
\def\aa{$\alpha$}
\def\coa{$^{56}$Co}
\def\cob{$^{57}$Co}
\def\nia{$^{56}$Ni}
\def\fea{$^{56}$Fe}
\def\feb{$^{57}$Fe}
\def\fec{$^{60}$Fe}
\def\nib{$^{57}$Ni}
\def\alb{$^{26}$Al}
\def\tii{$^{44}$Ti}
\def\sca{$^{44}$Sc}
\documentstyle[psfig]{article}
\begin{document}
\pagestyle{empty}
\hsize 5truein
\vsize 8truein
\font\abstract=cmr8
\font\keywords=cmr8
\font\caption=cmr8
\font\references=cmr8
\font\text=cmr10
\font\affiliation=cmssi10
\font\author=cmss10
\font\mc=cmss8
\font\title=cmssbx10 scaled\magstep2
\font\alcit=cmti7 scaled\magstephalf
\font\alcin=cmr6 
\font\ita=cmti8
\font\mma=cmr8
\def\ref{\par\noindent\hangindent 15pt}
\null


\title{\ni Gamma-Ray Line Astrophysics and Stellar  \\
Nucleosynthesis: Perspectives for INTEGRAL
}                                               
\bsk \bsk
\author{\ni N. Prantzos}
\bsk
\affiliation{ \ni Institut d'Astrophysique de Paris
}                                                
\bsk
\baselineskip = 12pt
\abstract{\ni ABSTRACT 
\ni
Nuclear \ga-ray lines constitute the most genuine diagnostic
tool of nuclear astrophysics, since they allow for an unambiguous
identification of isotopic species. 
Continuous improvement in instrumentation led to the discovery of several
radioactive species in the past 15 years (\alb, \coa, \cob, \tii) in
various astrophysical sites (SN1987A and SN1991T, Cas-A, the Milky Way
disk). These discoveries boosted theoretical activity on the nucleosynthesis
of these radioactive isotopes and on the refined modelling of the corresponding
sites, improving our knowledge on stellar evolution, stellar explosions,
galactic structure, etc.
I review here the current status of \ga-ray line astrophysics and present
some perspectives related to stellar nucleosynthesis, in view of future
missions like {\it INTEGRAL} (for more detailed recent reviews
see also Prantzos 1996; and Diehl and Timmes 1998).
}                                                    
\bsk
\baselineskip = 12pt
\bsk
\baselineskip = 12pt



\text{\ni 1. INTRODUCTION
\ssk
\ni     
The most important cosmic radioactivities for \ga-ray line astronomy
are presented in Table 1, along with the corresponding lifetimes,
energies, branching ratios, nucleosynthetic processes, astrophysical
sites of production and sites of detection (as of 1998).

The decay lifetime plays a determining role in the detectability
of a  radioactive nucleus. All nuclei are synthesized in high
density environments, initially opaque to \ga-rays. The photons
of their decays interact with the surrounding material and are
Compton-scattered down to X-ray energies, until they are
photoelectrically absorbed and their energy is released at longer
wavelengths. Only  in the case of a site violently
expanding after the explosion (SN, novae) or suffering extensive mass loss
(WR and AGB stars) do the \ga-ray lines have a chance to emerge, since
the opacity decreases with decreasing density.
 For instance, the timescale for the  supernova   ejecta to
 become transparent is such (a few weeks for SNIa, $\sim$1 yr for SNII)
 as to make essentially undetectable
 the \ga-ray lines of the short-lived \nia.

The intensity of the escaping \ga-ray lines gives important
information on the yields of the corresponding isotopes and the
physical conditions (temperature, density, neutron excess etc.) in the
stellar zones of their production. The shape of the \ga-ray lines reflects
the velocity distribution of the ejecta, modified by the opacity along
the line of site and can give information on the structure
of the ejecta (see e.g. Burrows 1991 for the potential of \ga-ray lines
as a tool of supernova diagnostics). Up to now, only the \coa \ lines
from SN1987A and the \alb \ line from the inner Galaxy have been resolved
(both with the same instrument, the  GRIS Ge spectrometer), but their
``message'' is not quite  understood yet.

When the lifetime of a  radioactive nucleus  is not very large w.r.t.
the timescale between two nucleosynthetic events in the Galaxy,
those events are expected to be seen as point-sources in the light
of that radioactivity. In the opposite case a diffuse galactic emission
is expected from the cumulated emission of many
sources.  Characteristic timescales between two
explosions  are $\sim$1-2 weeks for novae 
(from their estimated Galactic frequency of
$\sim$30 yr$^{-1}$, Della Vale and Livio 1994),
$\sim$40 yr for SNII+SNIb and $\sim$300 yr
for SNIa (from their corresponding Galactic frequencies,
Tammann et al. 1994).
Comparing those timescales to the decay lifetimes of Table 1
one sees that in the case of the long-lived \alb \ and \fec \
a diffuse emission is expected; the spatial profile of that emission
should reflect the Galactic distribution of the
underlying sources (except if high velocity ejecta can travel undecelerated
for long times; see Sec. 4).  The other radioactivities of Table 1
should be seen as point sources in the Galaxy except, perhaps,
$^{22}$Na from Galactic novae; indeed, O-Ne-Mg rich novae have a
frequency $\sim$1/3 of the total
resulting in $\sim$40 sources active  in the Galaxy during
the  lifetime of $^{22}$Na.

\def\mg{$^{26}$Mg}
\def\nia{$^{56}$Ni}
\def\nh{$^{60}$Ni}
\def\na{$^{22}$Na}
\def\co{$^{56}$Co}
\def\ci{$^{57}$Co}
\def\ch{$^{60}$Co}
\def\fe{$^{56}$Fe}
\def\fr{$^{57}$Fe}
\def\fh{$^{60}$Fe}
\def\ne{$^{22}$Ne}
\def\ti{$^{44}$Ti}
\def\sca{$^{44}$Sc}
\def\ca{$^{44}$Ca}
\def\be{$^{7}$Be}
\def\li{$^{7}$Li}
\def\ra{$\rightarrow$}
\def\aa{$\alpha$}
\def\nn{$\nu$}

\begin{table*}
{\small
\begin {center}
\caption{ {\bf TABLE 1:} COSMIC  RADIOACTIVITIES AND \ga-RAY LINES}
\begin{tabular}{ccccc}
\hline \hline
DECAY \ CHAIN &        MEAN \ LIFE$^*$          & LINE \ ENERGIES       &
 SITE  &  PROCESS \\
              & (yr) &  (MeV) & [Detected] & \\
\hline \noalign {\medskip}
\nia \ra \co \ra \fe & 0.31  & {\underline {0.847}} (1.) \
                   {\underline {1.238}} (0.685)  & SN  & NSE \\
  & & 2.598 (0.17) \ 1.771 (0.45)   & [SN1987A]    &    \\
                                        &       &    & [SN1991T]    &    \\
                                        &       &    &     &    \\
\ci \ra \fr &   1.1   & {\underline {0.122}} (0.86) \
                        {\underline {0.136}} (0.11)  & SN  & \aa-NSE  \\
            &       &    & [SN1987A]     &    \\
                                        &       &    &      &    \\
\na \ra \ne &   3.8   & 1.275 (1.)                   & Novae & $Ex.$H \\
                                        &       &    &      &    \\
\ti \ra \sca \ra \ca & 90 & {\underline {1.156}}(1.)
& SN & \aa-NSE \\
& & 0.068 (1.) \ 0.078 (0.98) & [CasA] &  \\
             &       &    &  &      \\
\alb \ra \mg &   1.1 \ 10$^6$   & {\underline {1.809}} (1.)
   & WR, AGB          &  $St.$H  \\
          &       &    &  Novae &   $Ex.$H        \\
          &       &    &  SNII  &  $St.+Ex.$Ne   \\
          &       &    & [Galaxy]             &  \nn  \\
          &       &    &               &       \\
\fh \ra \ch \ra \nh & 2.2 \ 10$^6$  &
1.322 (1.) \ 1.173(1.)  & SN   & n-NSE \\
\hline \hline
\end{tabular}
\end{center}
}
{\footnotesize *: For double decay chains the longest lifetime is given; \\
\ni $Underlined$: lines already detected; \\
\ni Numbers in $parentheses$: branching ratios;
In $brackets$: sites of lines detected \\
\ni $St.$: Hydrostatic burning; $Ex.$: Explosive burning;
NSE: Nuclear statistical equilibrium \\
\ni \aa: \aa-rich ``freeze-out'';  n-: normal ``freeze-out'';
\nn: neutrino-induced nucleosynthesis   }
\end{table*}

\bsk
\ni 2.   Co IN SN1987A, SN1991T ({\small AND EXTRAGALACTIC SNIa WITH {\it INTEGRAL}?} )
\ssk
\ni
The appearence of SN1987A in the closeby LMC confirmed in a spectacular way
early ideas about the synthesis of radioactive nuclei in supernovae;
in particular, it confirmed that \fea, the most strongly bound stable
nucleus in nature, is produced in the form of the unstable species \nia.
The main points of relevance to \ga-ray line astronomy are as follows:

1) The definitive confirmation of the \coa \ synthesis came from the
detection of the 0.847 MeV and 1.238 MeV lines  of its radioactive
decay (Matz et al. 1988). Their appearance $\sim$6
months earlier than expected suggested that mixing and/or fragmentation
had taken place in the mantle of SN1987A during the explosion or
shortly after, bringing heavy nuclei from the inner layers into the
outer ones. This discovery convincingly demonstrated  that the
explosion does not preserve the stratification of the onion-skin 
layers of the pre-supernova star. Prompted by this observation,
hydrodynamical 2-D and 3-D calculations     found that mixing does indeed take
place, due mostly  to  Rayleigh-Taylor type instabilities (e.g.
Hashisu et al. 1990, Fryxel et al.  1991).
That discovery  was a major contribution
of \ga-ray line astronomy to our understanding of supernova explosions.

2) The profile of the 0.847 MeV and 1.238 MeV lines, resolved by the
GRIS balloon-born instrument (Tueller et al. 1991) remains poorly
understood up to now. The lines are red-shifted by 500-800 km s$^{-1}$ (contrary
to what is expected from an optically thick source) and their
width is larger than predicted from theory
(indicating that some fraction of \coa \
has penetrated  deeply in the high velocity H-rich envelope). Despite some
preliminary models (Grant and Dean 1993;
Barrows and van Riper 1995), a convincing explanation does not exist yet.

3) The previous results were obtained before the CGRO launch.
The discovery by OSSE of the 122 keV and 136 keV lines  due to the
decay of \cob \ (Kurfess et al. 1992) is   the  major contribution of CGRO
in our understanding of SN1987A. The detected
flux corresponds to a mass of $\sim$2.7 10$^{-3}$ \ms \ of \cob.
This leads to a production ratio
\nib/\nia = 1.4$\pm$0.35 times the solar ratio of the daughter stable nuclei
(\feb/\fea)$_{\odot}$, suggesting an important  contribution from
``\aa-rich'' freeze-out, i.e. that most of \cob \ was produced in relatively
low-density environment (Clayton et al. 1992). Moreover, the OSSE
measurement ruled out earlier suggestions about the late lightcurve of SN1987A
being powered by a larger amount of \cob; recent
analysis of the latest UVBRIJHK lightcurves show that
the required ratio (57/56)$\sim$2 times solar is compatible
with \ga-ray measurements (Fransson and Kozma 1998).

The extraordinary chance offered by  SN1987A was due to its proximity; 
had it been in the
Andromeda galaxy, it would have been undetectable by instruments of the
80ies and 90ies and marginally detectable by an {\it INTEGRAL}-type instrument.
This is due to the fact that \ga-ray photons from massive star explosions
have to wait for a long time  before escaping, i.e. until the opacity of 
the $\sim$10 \ms \ slowly expanding envelope drops to sufficiently low 
levels. Much more interesting in that respect are SNIa:
their small envelopes
($<$1 \ms) and  large expansion velocities ($>$1.5 10$^4$ km/s), combined
to the large amounts of produced \nia \ ($\sim$0.5-1.  \ms, typically
ten times the average SNII yield), make their \coa \ lines thousands
of times brighter than those of SNII. The
 \coa \ lines of SNIa are detectable up to the Virgo cluster
of galaxies ($\sim$13-20 Mpc) by   instruments with a sensitivity
of $\sim$10$^{-5}$ \cs, i.e. close to the sensitivity limit of CGRO.

SN1991T, a bright SNIa,
exploded in the spiral galaxy NGC4527, at the periphery of the Virgo 
cluster and at an estimated  distance of 17 Mpc. 
An analysis of the COMPTEL data  for two
observations 66 and 176 days after the explosion, shows evidence
for the 847 and 1238 keV lines of \coa. 
The obtained line flux (Morris et al. 1995)
corresponds to a rather large amount of \nia ($>$1.3 \ms \
for a distance of $>$13 Mpc), implying that almost all of the
white dwarf turned into \nia. Sub-Chandrasekhar mass models for SNIa
(with a detonation in the base of  the accreted helium layer inducing
a further  detonation inside the white dwarf), or delayed detonation models
(where the flame front propagates subsonically at large distances from the
center of the white dwarf before turning into a detonation) may explain an
early detection of the \ga-ray lines, but perhaps not such large amounts
of \nia. More detections of extragalactic SNIa are required to clarify how
typical SN1991T was and before further conclusions are drawn.

SN1991T illustrates the kind of diagnosis of SNIa models that can be achieved
through the analysis of their \coa \ lines. A detailed exploration
of the potential of this method has been recently performed (H\"{o}flich et al.
1998, Gomez-Gomar et al. 1998). However, statistical analysis show that the
perspectives of detecting SNIa with {\it INTEGRAL} are rather dim, since
its sensitivity to broad lines (expected from the high velocity
ejecta of SNIa) is not
much better than that of CGRO (for recent estimates see Timmes and Woosley 1997;
and Isern, this volume).

\bsk
\ni 3.   $^{44}$Ti FROM CAS-A, GRO J0852-4642 ({\small AND SN1987A WITH {\it INTEGRAL} ?})
\ssk
\ni
The half-life of \tii \ has been surprisingly uncertain in the past 20 years,
with measured values ranging from 39 to 66 years. These discrepant results and
the difficulty in measuring the \tii \ lifetime are due to the fact that the
number of \tii \ nuclei one can obtain is small (w.r.t. the Avogadro number),
while the \tii \ lifetime is  large w.r.t. the available laboratory time.
The recent measurements of three different groups, however, seem to
convincingly converge towards a value of 60$\pm$1 years (G\" orres et al. 1998,
Ahmad et al. 1998, Norman et al. 1998).

The lifetime of \tii \ is comparable to the characteristic timescale
between two supernova explosions in the Milky Way, so that the
resulting \ga-ray line emission should appear as point sources in the
Galaxy. On the other hand, the \tii \ lifetime is sufficiently long to 
make   it an excellent probe of Galactic supernova explosions in the past
few centuries, since its \ga-ray lines may reveal supernova remnants
undetected in other wavelengths up to now (as has been demonstrated by
the recent detection of 1.16 MeV emission from the previously unknown
and presumably nearby SNR GRO J0852-4642; see Iyudin et al. 1998)

CasA is one of the youngest and closest known supernova remnants
at a distance of $\sim$3 kpc. Its age is evaluated to $>$300 yr from
extrapolation of the remnant's expansion to the origin. Despite its
proximity and high declination the explosion was not reported in
the 1600's. Optical and X-ray measurements suggest that the progenitor
was a 20 \ms \ WR star, that exploded as an underluminous SNIb.

The COMPTEL team reported the detection of the 1.16 MeV line of \tii,
with a flux of 4.2$\pm$0.9 10$^{-5}$ \cs (Iyudin et al. 1997)
compatible with the upper limit reported
by the OSSE team (5 10$^{-5}$ \cs, The et al. 1995). This
translates into a mass of \tii \ produced by the explosion of $\sim$1.5
10$^{-4}$ \ms \ i.e. not too far from theoretical predictions.

However, the CasA detection brought forward an unexpected puzzle. Since both 
\tii \ and \nia \ are synthesized in the same stellar zones,
the inferred amount of \tii \
corresponds to  a mass of \nia \
of $\sim$0.05 \ms. Powered by the \nia \ decay, the supernova (with or without
a hydrogen envelope) would have then a peak magnitude
M$_V <$-4 at 3 kpc (The et al. 1995).   CasA should have been a rather bright
supernova for a few weeks, making it difficult to understand why it
went unreported. Some 10 magnitudes of visual extinction are required to make
the \tii \ observation consistent with the absence of historical records
for that supernova.

Among the various proposed solutions to the puzzle, 
the most natural seems to be the one invoking a dusty shell of material
(ejected by the winds of the progenitor star) hiding CasA during the
explosion; the supernova shock wave would have destroyed later most of the
dust, as it propagated through the debris, explaining the currently
dust-free environment of CasA (Hartmann et al. 1997).

Notice that \tii \ is the third radioactivity that will, someday, be discovered in
SN1987A. Along with \cob \ it is produced in the hottest and deepest
layers ejected during the explosion of a massive star.
The ejected quantity is very sensitive to the position of the ``mass-cut'' (i.e.
the line dividing the supernova ejecta from the matter 
accreted onto the compact object).
Nucleosynthesis calculations for SN1987A show that
$\sim$1.5 10$^{-4}$ \ms \ of \tii \ were synthesized  by the explosion
(e.g. Thielemann et al. 1996); similar amounts  are  suggested from the
fitting of the late lightcurve of SN1987A, if it  is indeed powered by \tii \
(Fransson and Kozma 1997). 
Such amounts of \tii \ produce on Earth a flux of $\sim$3 10$^{-6}$ \cs \ 
in the 68, 78 and 1157 keV lines
for many decades after the explosion. If \tii \ is ejected at low speed
($\sim$10$^3$ km/s, as suggested by spherically symetric models of the
explosion of SN1987A), the kinetic broadening of the lines will be small
and the estimated fluxes close to the sensitivity
limit of {\it INTEGRAL} for narrow lines.  \tii \ from SN1987A
will then certainly be a prime target
for that satellite; its detection will allow to probe the deepest
ejected layers of SN1987A better than ever before.

\bsk
\ni 4.    \alb \ IN THE GALACTIC PLANE  ({\small AND \fec \ WITH {\it INTEGRAL}?})
\ssk
\ni
\alb \ is the first cosmic radioactivity ever detected in \ga-rays, through
its characteristic 1.8 MeV line. Its detection by the HEAO-3 satellite
(Mahoney et al. 1982) clearly shows that nucleosynthesis is currently
active in the Galaxy  and offers an unprecedented opportunity
to identify the site of that activity.
A recent review (Prantzos and Diehl 1996)
discusses all  aspects of \alb \ relevant to nucleosynthesis
and \ga-ray line astronomy.

The current status of \alb \
observations by COMPTEL is presented by Diehl (this volume).
The data shows clearly a diffuse, irregular, emission  along the
Galactic plane, allowing to eliminate: i) a unique point source
in the Galactic centre;
ii) a strong contribution of the Galactic bulge, signature of an
old population and iii) any class of sources involving a large number
of sites with low individual yields (novae, low mass AGB stars),
since a smooth flux distribution is expected in that case.

The irregular 1.8 MeV emission  detected by COMPTEL 
along the galactic plane reveals, better than any other tracer, the sites of
current nucleosynthetic activity in the Galaxy. A very tempting  identification 
can already be made of several 1.8 MeV "hotspots" with tangents to the
spiral arms (Diehl et al. 1995); as suggested by Prantzos (1993)
such a correlation  would imply that massive stars are at the origin of
(most of) the derived $\sim$2.5 \ms \ of galactic \alb.

In a recent work of multi-frequency image comparison, Kn\"{o}dlseder (1998)
showed that a map of the ionisation power from massive stars (derived from
the COBE data, after correction for synchrotron contribution) corresponds
to the 1.809 MeV map of galactic \alb \ in all significant detail; assuming a
standard stellar initial mass function, his calculation reproduces consistently
the current galactic supernova rate and massive star population from both maps,
and suggests that most of \alb \ is produced by WR stars of high metallicity in the
inner Galaxy.

Two of the  COMPTEL hotspots, at l$\sim$80$^o$ and  90$^o$ are certainly
not related to spiral features (Cygnus superbubble, Vela region).
Their detection  will certaibly allow to probe better
the underlying stellar populations  and their \alb \ yields; notice that
the 1.8 MeV Vela "hot-spot" is no more associated to the closeby 
Vela supernova remnant alone, as originally  thought (Diehl 1998, this
volume).

A recent intriguing development in the \alb \ ``saga'' concerns the spectral
width of the 1.8 MeV emission from the inner Galaxy.
The {\it GRIS} spectrometer resolved the 1.8 MeV line from that
region (Naya et al. 1996), finding it larger ($\Delta$E=5.4$\pm$1.4 keV keV)
than what expected from galactic rotation ($\sim$1 keV).
Even if \alb \ is initially ejected at high velocities, it is difficult at
present to understand how it could go undecelarated during most of its 1 Myr
lifetime. Among the several alternative hypotheses explored  in Chen et al. (1997)
the one of \alb \ being condensed in high-speed dust grains seems promising,
but the {\it GRIS} measurement needs conformation, since it is incompatible with
the {\it HEAO C} line width limit of $<$3 keV.

One of the most exciting perspectives for {\it INTEGRAL} in relation to \alb,
 is the possibility to detect a diffuse $^{60}$Fe $\gamma$-ray emission at 
1.2 and 1.3 MeV  and give some hints as to its distribution in the Galaxy.
Indeed, detailed SNII models (Woosley and Weaver 1995)
give a \fec \ yield $\sim$0.25-0.35 of the
corresponding \alb \ yield  when
averaged over a reasonable stellar initial mass function. Taking into
account their respective lifetimes (Table 1), the flux in the \fec \
lines is expected to be $\sim$0.15 that of the galactic 1.8 MeV flux,
 {\it if} SNII are the major producers of Galactic \alb. If
the \fec \ lines are not detected by {\it INTEGRAL}, then nucleosynthesis in
SNII  should be seriously revised or (more radically) SNII
should be discarded as major sources of \alb.  In both cases a major
information for stellar nucleosynthesis will be obtained.
\newpage
\def\ms{M$_{\odot}$}
\def\zs{Z$_{\odot}$}
\def\co{$^{56}$Co}
\def\nia{$^{56}$Ni}
\def\ti{$^{44}$Ti}
\def\fe{$^{60}$Fe}
\def\ga{$\gamma$}
\def\aa{$\alpha$}
\def\lr{$\rightarrow$}
\def\De{$\Delta E=\pm$}
\def\pcs{cm$^{-2}$ s$^{-1}$}

\begin{table*}
\begin{center}
{\small
\caption{ {\bf TABLE 2: }PERSPECTIVES FOR STELLAR \ga-RAY LINES WITH {\it INTEGRAL}}
\begin{tabular}{ccccc}
\hline \hline
ISOTOPE    & LINE E(MeV) & TARGET    & OBSERVABLE &  INTEREST \\
\hline
\hline \noalign {\smallskip}
 {\bf \co} & 0.847  & Extragalactic & Intensity  & Constrain models  \\
           & 1.238  & SNIa    & Shape      &  of SNIa          \\
\noalign {\smallskip}
\hline \noalign       {\smallskip}
          &   &  SN1987A    & Flux & Nucleosynthesis    \\
          &   &             &      & Mass-cut     \\
          & 0.068   &             &                         &             \\
{\bf \ti} & 0.078  &  CasA       & Confirmation            & V$_{EJ}$\lr CasA Age   \\
          & 1.156  &             & + Shape                 &  + Models    \\
          &   &             &                         &             \\
          &   & Galactic SN &  Intensity           &  Models +   \\
          &   &             &   + Shape            & Nucleosynthesis \\
\noalign  {\smallskip}
\hline \noalign       {\smallskip}
          &        &  Galaxy       & Accurate map     & Nucleosynthesis sites \\
          &         &              & Line width       &  Ejecta propagation   \\
{\bf \alb}&1.809   &              &                  &                        \\
          &   & Galactic          & Line Shape       & Distances          \\
          &   & ``Hot-spots''      & Flux            & Yields          \\
          &   &  (e.g. Vela )     & Extent           & V$_{EJ}$          \\
\noalign    {\smallskip}
\hline \noalign        {\smallskip}
  {\bf \fe}& 1.173 & Galaxy   & F$\sim$ 0.1 F$_{^{26}Al}$ & Sources    \\
           & 1.322  &               &  (map ?)        &                        \\
\noalign     {\smallskip}
\hline \noalign          {\smallskip}
  {\bf \na}& 1.275 & Novae    & Flux, Shape           & Models     \\
\noalign {\smallskip}
\hline \hline
\end{tabular}
}
\end{center}
\end{table*}

\bsk
\ni 5.    SUMMARY AND PERSPECTIVES
\ssk
\ni
Gamma-ray line astronomy became a privileged tool for the study of
stellar nucleosynthesis in the past ten years; the proximity of
SN1987A and the launch of CGRO played a major role in this rapid
progress.

The detection of four cosmic radioactivities up to now (\coa,
\cob, \tii, and \alb) allowed to probe in depth the  supernova
structure and the thermodynamic conditions of the explosion
(with \coa \ and \cob \ in SN1987A), 
to confront simple supernova models
to observations (with more success in SN1991T than in CasA)
and to locate the sites of large scale nucleosynthetic activity in the
Galaxy (through the irregular  profile of the \alb \ emission).

The perspectives are even brighter, since
the increased sensitivity of {\it INTEGRAL} will allow to explore further
those issues and to tackle several other related topics. 
A synopsis of the most
important  perspectives for the study of stellar
nucleosynthesis with {\it INTEGRAL} is presented in Table 2.
Most of those perspectives are discussed in the previous sections.
For those left outside this review, one should see
the contributions by Hernanz et al. (for $^{22}$Na from galactic novae)
and Leising (for the galactic 511 keV line), in this volume.

 }

\bsk
\baselineskip = 12pt
{\abstract \ni ACKNOWLEDGMENTS
I am indebted to R. Diehl, J. Kn\"{o}dlseder and M. Leising for many stimulating
discussions on \ga-ray line astronomy and to the organisers for their kind invitation
in this workshop.}

\bsk
\baselineskip = 12pt


{\references \ni REFERENCES
\ssk

\ref  Ahmad  F., et al., 1998, Nucl. Phys., submitted
\ref  Burrows A., 1991, in ``Gamma-Ray Line Astrophysics'', Eds.
            Ph. Durouchoux \& N. Prantzos (AIP Conf. Proc. 232), p. 297
\ref  Burrows A. \& van Riper K.,  1995, ApJ, 455, 215
\ref  Clayton D.,  Leising M., The L.S., Johnson W. \& Kurfess J.,
            1992, ApJ, 399, L141
\ref  Della Valle M \& Livio M., 1994, A\&A, 287, 403
\ref  Diehl R. et al., 1995, A\&A, 298, 445
\ref  Diehl R. and Timmes F., 1998, PASP, 110, 637
\ref  Fransson C. \& Kozma C., 1998, in ``SN1987A: 10 Years After'',
      Eds. M. Philips and N. Suntzeff (ASP), in press
\ref  Fryxel B., Muller E. \& Arnett D., 1991, ApJ, 367, 619
\ref  Gomez-Gomar J., Isern J. \& Jean P., 1998, MNRAS, 295, 1
\ref  G\" orres J., et al., 1998, Phys. Rev. Lett., 80, 2554
\ref  Grant K. \& Dean  A., 1993, A\&A Suppl., 97, 211
\ref  Hartmann D. et al. 1997, Nucl. Phys. A621, 83
\ref  H\"{o}flich P., Wheeler J.C. \& Khokhlov A. 1998, ApJ, 492, 228
\ref  Hashisu I., Matsuda T., Nomoto K. \& Shigeyama T., 1990,
            ApJ, 358, L57
\ref  Iyudin A., et al. 1997, ``The Transparent Universe'', Eds. C.
      Winkler et al., (ESA-SP 382),  37
\ref  Iyudin A., et al. 1998, Nature, 396, 142
\ref  Kn\"{o}dlseder J., 1998, ApJ, in press
\ref  Kurfess J. et al., 1992, ApJ, 399, L137
\ref  Mahoney W., Ling J., Jacobson A. \& Lingelfelter R., 1982
            ApJ, 262, 742
\ref  Matz S. et al., 1988, Nature, 331, 416
\ref  Morris D. et al., 1995, in 17th Texas Symp. on Relat. Astr.,
            (New York), p.397
\ref  Naya J. et al., 1996, Nature, 384, 44
\ref  Norman E. et al., 1998, Phys Rev C, submitted
\ref  Prantzos N., 1993, ApJ, 405, L55
\ref  Prantzos N., 1996, A\&A Suppl., 120, 303
\ref  Prantzos N. \& Diehl R., 1996, Phys. Rep., 267, 1
\ref  Tammann G., Loffler W. \& Schroder A., 1994, ApJ Suppl.,
            92, 487
\ref  The L.-S. et al.,  1995, ApJ, 444, 244
\ref  Thielemann F.-K., Nomoto K. \& Hashimoto M., 1996, ApJ, 460, 408
\ref  Timmes F. \& Woosley S., 1998, ApJ, 489, 160
\ref  Tueller J., Barthelmy S., Gehrels N., Teegarden B.,
            Leventhal M. \& MacCallum C., 1991, ApJ, 351, L41
\ref  Woosley S. \& Weaver T., 1995, ApJS, 101, 181
            300  }

\end{document}